\documentclass{ws-mpla}

\begin{document}

\markboth{Cai-Dian L\"u} {QCD in hadronic B decays}

%
\catchline{}{}{}{}{}
%

\title{QCD in hadronic B decays
\footnote{work supported by national science foundation of China.}
}

\author{\footnotesize Cai-Dian L\"u}

\address{Institute of High Energy Physics, CAS, P.O. Box
918(4), Beijing 100049, China\\
lucd@ihep.ac.cn}

\maketitle

\pub{Received (10 March 2005)} { }

\begin{abstract}
The perturbative QCD approach is based on $k_T$ factorization,
including the Sudakov form factors so that to avoid the endpoint
singularity.  In this approach, we calculate the charmless B
decays like $B\to \pi\pi$ decays etc. to produce the right number
of branching ratios and also CP asymmetry parameters. For final
states with at least one charmed meson, like $B\to D\pi$ decays,
our results also agree with the experiments.

\end{abstract}

\ccode{PACS Nos.: include PACS Nos.}

\section{Introduction}

The hadronic B decays are important for the CKM matrix elements
measurements and CP violation detection. Understanding nonleptonic
$B$ meson decays is crucial for testing the standard model, and
also for uncovering the signal of new physics. The simplest case
is two-body hadronic $B$ meson decays, for which Bauer, Stech and
Wirbel (BSW) proposed the naive factorization assumption (FA) in
their pioneering work \cite{BSW}. Considerable progress, including
generalized FA \cite{akl,Cheng94} and QCD-improved FA (QCDF)
\cite{BBNS}, has been made since this proposal. On the other hand,
technique to analyze hard exclusive hadronic scattering was
developed by Brodsky and Lepage \cite{LB} based on collinear
factorization theorem in perturbative QCD (PQCD). A modified
framework based on $k_T$ factorization theorem was then given in
\cite{BS,LS}, and extended to exclusive $B$ meson decays in
\cite{LY1,li,CLY}. The infrared finiteness and gauge invariance of
$k_T$ factorization theorem was shown explicitly in \cite{NL}.
Using this so-called PQCD approach, we have investigated dynamics
of nonleptonic $B$ meson decays \cite{KLS,LUY,KS}.

Although the predictions of branching ratios agree well with
experiments in most cases, there are still some theoretical points
unclear in FA and QCDF. First, it relies strongly  on the form
factors, which cannot be calculated by FA itself. Secondly, the
generalized FA shows that the non-factorizable contributions are
important in a group of channels, which can not be done reliably
in FA and QCDF \cite{akl,Cheng94}. The reason of this large
non-factorizable contribution needs more theoretical studies.
Thirdly, the strong phase, which is important for the CP violation
prediction, is quite sensitive to
 various approaches. The mechanism of this strong phase is quite
 different for different method, and give quite different results.
The recent experimental results can make a test for the validity
of these approaches.

\section{Formalism of PQCD Approach}

In this section, we will introduce the idea of PQCD  approach. The
three scale PQCD factorization theorem has been developed for
non-leptonic heavy meson decays \cite{li}, based on the formalism
by Brodsky and Lepage \cite{LB}, and Botts and Sterman \cite{BS}.
In the non-leptonic two body B
 decays, the $B$ meson is heavy, sitting at rest.
It decays into two light mesons with large momenta. Therefore the
light mesons are moving very fast in the rest frame of $B$ meson.
In this case, the short distance hard dynamic dominates the decay
amplitude. The reasons can be ordered as: first, because there are
not many resonance near the energy region of $B$ mass, so it is
reasonable to assume that final state interaction is not important
in two-body $B$ decays. Second, With the final light mesons moving
very fast, there must be a hard gluon to kick the light spectator
quark (with small momentum) in the B meson to form a fast moving
light
 meson. So the dominant diagram in this theoretical picture
is that one hard gluon from the spectator quark connecting with
the other quarks in the four quark operator of the weak
interaction. Unlike the usual factorization approach, the hard
part of the PQCD approach consists of six quarks rather than four.
We thus call it six-quark operators or six-quark effective theory.
There are also soft (soft and collinear) gluon exchanges between
quarks. Summing over those leading soft contributions gives a
Sudakov form factor, which suppresses the soft contribution to be
dominant. Therefore, it makes the PQCD reliable in calculating the
non-leptonic decays. With the Sudakov resummation, we can include
the leading double logarithms for all loop diagrams, in
association with the soft contribution.


There are three different scales in the B meson non-leptonic
decay. The QCD corrections to the four quark operators are usually
summed by the renormalization group equation \cite{buras}. This
has already been done to the leading logarithm and next-to-leading
order for years. Since the $b$ quark decay scale $m_b$ is much
smaller than the electroweak scale $m_W$, the QCD corrections are
non-negligible. The third scale $1/b$ involved in the $B$ meson
exclusive decays is usually called the factorization scale, with
$b$ the conjugate variable of parton transverse momenta. The
dynamics below $1/b$
 scale is regarded as being completely
non-perturbative, and can be parameterized into meson wave
functions. The meson wave functions are not calculable in PQCD.
But they are universal, channel independent. We can determine them
 from experiments, and it is constrained   by QCD sum rules and
Lattice QCD calculations. Above the scale $1/b$, the physics is
channel dependent. We can use perturbation theory to calculate
channel by channel.

Besides the hard gluon exchange with the spectator quark, the soft
gluon exchanges between quark lines  give out the double
logarithms $\ln^2(Pb)$ from the overlap of collinear and soft
divergence, $P$ being the dominant light-cone component of a meson
momentum. The resummation of these double logarithms leads to a
Sudakov form factor $\exp[-s(P,b)]$, which suppresses the long
distance contributions in the large $b$ region, and vanishes as
$b> 1/\Lambda_{QCD}$.   So this term includes the double infrared
logarithms. The corresponding Sudakov factor can be derived in
PQCD as a function of the transverse separation $b$ and of the
momentum fraction $x$ carried by the spectator quark \cite{LY1}.
The Sudakov factor suppresses the large $b$ region, where the
quark and antiquark are separated by a large transverse distance
and the color shielding is not so effective. It also suppresses
the $x\sim 1$ region, where a quark carries all of the meson
momentum, and intends to emit real gluons in hard scattering. The
Sudakov factors from $k_T$ resummation \cite{CS} for the $B$ and
$D^{(*)}$ mesons are only associated with the light spectator
quarks, since the double logarithms arise from the overlap of the
soft and mass (collinear) divergences.
 It is  shown in ref.\cite{LUY} that $e^{-s}$ falls
off quickly in the large $b$, or long-distance, region, giving
so-called Sudakov suppression. This Sudakov factor practically
makes PQCD approach applicable. For the detailed derivation of the
Sudakov form factors, see ref.\cite{LY1,8}.

With all the large logarithms resummed, the remaining finite
contributions are absorbed into a perturbative b quark decay
subamplitude $H(t)$. Therefore the three scale factorization
formula is given by the typical expression,
\begin{equation}
C(t) \times H(t) \times \Phi (x) \times \exp\left[ -s(P,b) -2 \int
_{1/b}^t \frac{ d \bar\mu}{\bar \mu} \gamma_q (\alpha_s (\bar
\mu)) \right], \label{eq:factorization_formula}
\end{equation}
where $C(t)$ are the corresponding Wilson coefficients, $\Phi (x)$
are the  meson wave functions and the variable $t$ denotes the
largest mass scale of hard process $H$, that is, six-quark
effective theory.
 The quark anomalous dimension $\gamma_q=-\alpha_s /\pi$ describes
the evolution from scale $t$ to $1/b$. Since logarithm corrections
have been summed by renormalization group equations, the above
factorization formula does not depend on the renormalization scale
$\mu$ explicitly.

        \begin{figure}[tbp]
\epsfig{file=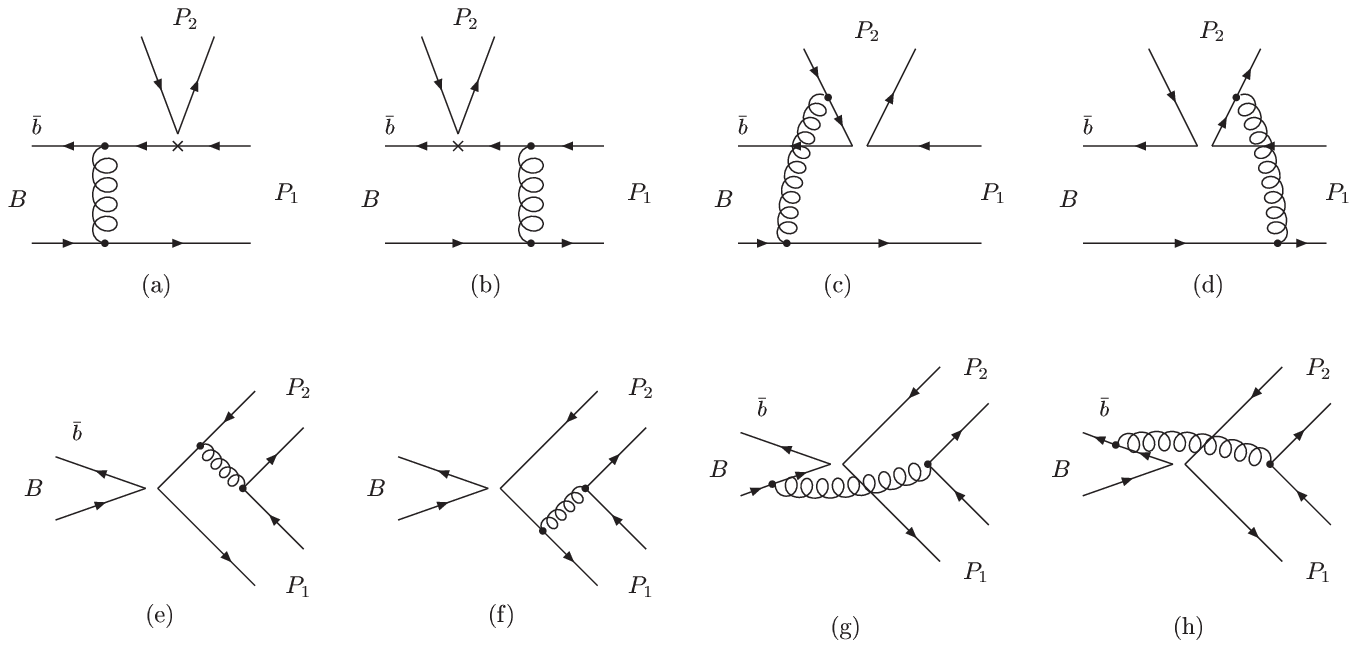,bbllx=4.6cm,bblly=18.9cm,bburx=15.5cm,bbury=25cm,%
width=9.6cm,angle=0}
    \caption{Diagrams for $B\to P_1P_2$ decay in perturbative QCD approach. The factorizable
    diagrams (a),(b), non-factorizable (c),
    (d),  factorizable annihilation
    diagrams (e),(f) and non-factorizable annihilation diagrams (g),(h).}
    \label{fig2}
   \end{figure}

As shown above, in the PQCD approach, we keep the $k_T$ dependence
of the wave function. In fact, the approximation of neglecting the
transverse momentum can only be done at the non-endpoint region,
since $k_T \ll k^+$ is qualified  at that region. At the endpoint,
$k^+ \to 0$, $k_T$ is not small any longer, neglecting $k_T$ is a
very bad approximation. By, keeping the $k_T$ dependence,
 there is no endpoint divergence as
occurred in the QCD factorization approach, while the numerical
result does not change at other region. Furthermore, the Sudakov
form factors suppress the endpoint region of the wave functions.
Recently another type of resummation has been observed. The loop
correction to the weak decay vertex produces the double logarithms
$\alpha_s\ln ^2 x_2$ \cite{threshold}. Using the wave functions
 from light-cone sum rules, at the endpoint region, these large
logarithms are important, they must be resummed. The threshold
resummation for the jet function results in Sudakov suppression,
which decreases the contribution of endpoint region of wave
functions. Therefore, the main contributions to the decay
amplitude in PQCD approach comes not from the endpoint region. The
perturbative QCD is applied safely.

The main input parameters in PQCD are the meson wave functions. It
is not a surprise that the final results are   sensitive to the
meson wave functions. Fortunately, there are many channels involve
the same meson, and the meson wave functions should be process
independent. In all the calculations of PQCD approach, we follow
the rule, and we find that   they can explain most of the measured
branching ratios of B decays.   For example: $B\to \pi\pi$ decays,
$B\to \pi \rho$,
  $B\to \pi \omega$ decays \cite{LUY}, $B\to K\pi$ decays \cite{KLS}, $B\to K K$ decays
  \cite{bkk}, the form factor calculations of $B\to \pi$, $B\to \rho$
  \cite{semi}, $B\to K \eta^{(\prime)}$ decays \cite{eta},
   $B\to K\phi$ decays \cite{bkphi}    etc.

We emphasize that nonfactorizable and annihilation diagrams are
indeed subleading in the PQCD formalism as $M_B\to \infty$. This
can be easily observed from the hard functions in appendices of
ref.\cite{KLS,LUY}. When $M_B$ increases, the $B$ meson wave
function enhances contributions to factorizable diagrams. However,
annihilation amplitudes, being independent of B meson wave
function, are relatively insensitive to the variation of $M_B$.
Hence, factorizable contributions become dominant and annihilation
contributions are subleading in the $M_B\to\infty$ limit
\cite{bkphi}. Although the non-factorizable and annihilation
diagrams are subleading for the branching ratio in color enhanced
decays, they provide the main source of strong phase, by inner
quark or gluon on mass shell. This mechanism of strong phase is
negligible in the PQCD approach, since it is at next-to-leading
order $O(\alpha_s)$ corrections. In fact, the factorizable
annihilation diagrams are Chirally enhanced. They are not
negligible in PQCD approach \cite{KLS,LUY}.

 In the PQCD framework based on $k_T$ factorization theorem,
an amplitude is expressed as the convolution of hard $b$ quark
decay kernels with meson wave functions in both the longitudinal
momentum fractions and the transverse momenta of partons. In the
$B\to D\pi$ like decays with at least one heavy meson in the final
states, our PQCD formulas are derived up to leading-order in
$\alpha_s$, to leading power in $m_D/m_B$ and in
$\bar\Lambda/m_D$, and to leading double-logarithm resummations.

\section{Numerical Results and Discussion}

The PQCD predictions for each term of the $B\to D\pi$ decay
amplitudes \cite{dpi} are exhibited in Table~\ref{dp}. The
theoretical uncertainty comes only from the variation of the shape
parameter for the $D$ meson distribution amplitude, $0.6 < C_D <
1.0$. It is expected that the color-allowed factorizable amplitude
$f_\pi\xi_{\rm ext}$ dominates, and that the color-suppressed
factorizable contribution $f_D\xi_{\rm int}$ is smaller due to the
Wilson coefficient $C_1+C_2/N_c\sim 0$. The color-allowed
non-factorizable amplitude ${\cal M}_{\rm ext}$ is negligible:
since the pion distribution amplitude is symmetric under the
exchange of $x_3$ and $1-x_3$, the contributions from the two
diagrams Figs.~\ref{fig2}(c) and \ref{fig2}(d) cancel each other
in the dominant region with small $x_2$. It is also down by the
small Wilson coefficient $C_1/N_c$. For the color-suppressed
non-factorizable contribution ${\cal M}_{\rm int}$, the above
cancellation does not exist in the dominant region with small
$x_3$, because the $D$ meson distribution amplitude $\phi_D(x_2)$
is not symmetric. Furthermore, ${\cal M}_{\rm int}$, proportional
to $C_2/N_c\sim 0.3$, is not down by the Wilson coefficient. It is
indeed comparable to the color-allowed factorizable amplitude
$f_\pi\xi_{\rm ext}$, and produces a large strong phase. Both the
factorizable and non-factorizable annihilation contributions are
vanishingly small.

\begin{table}[t]
\tbl{Predicted $B\to D\pi$ decay amplitudes in units of $10^{-2}$
GeV, and branching ratios in units of
$10^{-3}$.}{\begin{tabular}{|l| c| c|c|c| }
\hline Quantities & $C_D=0.6$ & $C_D=0.8$ & $C_D=1.0$ & Data \\
\hline
${\cal M}_1$ &$6.39-0.35i$ &$6.88-0.38i$ &$7.35-0.40i$ &\\
${\cal M}_2$ &$-1.53+1.48i$ &$-1.49+1.48i$ &$-1.45+1.45i$ &\\
${\cal M}_3$ &$8.56-2.45i$ &$8.99-2.47i$ &$9.40-2.46i$ &\\
\hline
$B(\bar B^0\to D^+\pi^-)$ & 2.37 & 2.74 & 3.13 & $3.0\pm 0.4$ \\
$B(\bar B^0\to D^0\pi^0)$ & 0.26 & 0.25 & 0.24 & $0.29 \pm 0.05$\\
$B(B^-\to D^0\pi^-)$ & 4.96 & 5.43 & 5.91 & $5.3\pm 0.5$ \\
\hline
\end{tabular}}
\label{dp}
\end{table}

The predicted branching ratios in Table~\ref{dp} are in agreement
with the averaged experimental data \cite{BelleC}. We extract the
parameters $a_1$ and $a_2$ from our calculations of amplitudes.
That is, our $a_1$ and $a_2$ do not only contain the
non-factorizable amplitudes as in generalized FA, but the small
annihilation amplitudes, which was first discussed in \cite{GKKP}.
We obtain the ratio $|a_2/a_1|\sim 0.43$ with $10\%$ uncertainty
and the phase of $a_2$ relative to $a_1$ about $Arg(a_2/a_1)\sim
-42^\circ$.   Note that the experimental data do not fix the sign
of the relative phases. The PQCD calculation indicates that
$Arg(a_2/a_1)$ should be located in the fourth quadrant. It is
evident that the short-distance strong phase from the
color-suppressed nonfactorizable amplitude is already sufficient
to account for the isospin triangle formed by the $B\to D\pi$
modes.   Hence, it is more reasonable to claim that the data just
imply a large strong phase, but do not tell what mechanism
generates this phase \cite{JPL}. From the viewpoint of PQCD, this
strong phase is of short distance, and produced from the
non-pinched singularity of the hard kernel. Certainly, under the
current experimental and theoretical uncertainties, there is still
room for long-distance phases from final-state interaction. Other
decays with one $D$ meson in final states, like $B\to D_s^{(*)}
K^{(*)}$ \cite{dsk}, $B\to D^{(*)}\eta^{(\prime)}$  and $B\to D_s
\phi$ \cite{deta} etc. also agree with experiments.

\begin{table}[t]
\tbl{Direct CP asymmetries calculated in FA, QCDF and PQCD for
$B\to \pi\pi$ and $B\to K\pi$ decays together with the
experimental results at percentage.}{\begin{tabular}{|l| c| c|c|c|
}
\hline Quantities & FA & QCDF & PQCD & Data \\
\hline
$ B^0\to \pi^+\pi^-$ &      $-5\pm 3$   &    $-6\pm 12$ &     $+30\pm 20$ &    $+37\pm 11$ \\
$ B^0\to \pi^+K^-$ &$+10\pm 3$  &$+5\pm 9$  & $-17\pm 5$ &$-10.9\pm 1.9$\\
$B^+\to K^0\pi^+$ & $+1.7\pm 0.1$  &$ +1 \pm 1$ &$  -1.0\pm 0.5$   &$-2.0 \pm 3.4$ \\
$B^+\to K^+\pi^0$ & $ +8\pm 2$  &$+ 7 \pm 9$ &$  -13\pm 4$   &$ +4 \pm 4$ \\
\hline
\end{tabular}}
\label{cp}
\end{table}

As discussed in the previous section, the strong phase generated
 from PQCD approach is quite different from the FA and QCDF
approaches. The direct CP asymmetry is proportional to the sine of
the strong phase difference of two amplitudes. Therefore the
direct CP asymmetry will be different if strong phase is
different. The predicted CP asymmetry by the three methods are
shown in table \ref{cp}. It is easy to see that the FA \cite{akl}
and QCDF \cite{BBNS} results are quite close to each other, since
the mechanism of strong phase is the same for them. Recently the
two B factories measure at least two channels with non-zero direct
CP asymmetry: $B^0\to \pi^+\pi^-$ and $B^0\to \pi^+K^-$
\cite{cpa}, which are shown in table \ref{cp}. It is easy to see
that our PQCD results of direct CP asymmetry \cite{LUY} agree with
the experiments. Although FA and QCDF are not yet ruled out by
experiments, but the experiments at least tell us that the
dominant strong phase should come from the mechanism of PQCD not
QCDF. Charm quark loop mechanism, which gives the central value of
strong phase in QCDF, is argued next-to leading order in PQCD.
This argument is now proved by B factories experiments.

\section{Summary}

In the PQCD approach, the form factors are  calculable, which are
 dominant by short distance contribution. By including the $k_T$ dependence and
Sudakov suppression, there is no endpoint divergence. In the PQCD
formalism non-factorizable  amplitudes are of the same order as
factorizable ones in powers of $1/M_B$, which are both
$O(1/(M_B\Lambda_{\rm QCD}))$.
 The smaller magnitude of nonfactorizable amplitudes in color
 enhanced decays are
due to the cancellation of the two non-factorizable diagrams. From
the viewpoint of power counting, they are of the same order. In
case of $B\to D \pi$ decays, the cancellation is absent. The power
counting changes, so that we   can also calculate the
non-factorizable contribution dominant processes. In this case, we
give the right branching ratios for $B\to D^0 \pi^0$ decay.

The strong phase comes mainly from the annihilation and
non-factorizable diagrams in PQCD approach, which is quite
different from the FA and QCDF approaches. The experimentally
measured direct CP asymmetry implies that PQCD gives at least the
dominant strong phase than other approaches. This will be further
tested by experiments.

\end{document}